\documentclass[12pt,preprint]{aastex}

\slugcomment{Accepted for publication in AJ}

\shorttitle{Physical Properties of Gliese 229B}
\shortauthors{Nakajima et al.}

\begin{document}


\title{Physical Properties of Gliese 229B
    Based on Newly Determined Carbon and Oxygen Abundances of Gliese 229A}


\author{T. Nakajima}
\affil{National Astronomical Observatory of Japan, 2-21-1, Osawa, Mitaka, Tokyo, 181-8588, Japan}
\email{tadashi.nakajima@nao.ac.jp}

\author{T. Tsuji} 
\affil{Institute of Astronomy, School of Science, The University of Tokyo, 2-21-1 Osawa, Mitaka, Tokyo, 181-0015, Japan}
\email{ttsuji@ioa.s.u-tokyo.ac.jp}

\and

\author{Y. Takeda}
\affil{National Astronomical Observatory of Japan, 2-21-1, Osawa, Mitaka, Tokyo, 181-8588, Japan}
\email{takeda.yoichi@nao.ac.jp}




\begin{abstract}

Recently \citet{tsu14} and \citet{tsu15} have developed a method of molecular line spectroscopy of M dwarfs
with which carbon and oxygen abundances are derived respectively from CO and H$_2$O lines in the $K$ band.
They applied this method to Gliese 229A(Gl 229A), the primary star of the brown dwarf companion, Gliese 229B(Gl 229B).
The derived abundances of Gl 229A are
$\log A_{\rm C}=-3.27\pm0.07$ and $\log A_{\rm O}=-3.10\pm0.02$, which are close to the classical values of
the solar abundances of carbon and oxygen. 
We generate model spectra of Gl 229B for the metallicity of Gl 229A as well as for the classical solar metallicity
($\log A_{\rm C} = -3.40$ and $\log A_{\rm O} = -3.08$).
We find that the differences of the resulting spectra are not so large for the differences of the metallicity
of 0.1 dex or so, but we now discuss the spectrum of Gl 229B on the basis of the reliable metallicity.
From the literature \citep{giz02}, the lower limit to the age of Gl 229A is found to be 0.3 Gyr.
From the kinematics of Gl 229A, we evaluate the upper limit  to the age of Gl 229A
to be 3.0 Gyr.
The observed and model spectra are compared and goodness of fit is obtained in the range of model parameters,
$750\leq T_{\rm eff}\le1000$K, and $4.5\leq\log g\le5.5$.
Among the candidates that satisfy the age constraint, 
the best combinations of model parameters are 
$(T_{\rm eff},\log g)$=(800K,4.75) and (850K,5.0),
while acceptable combinations are (750K,4.75),(850K,4.75) and (900K,5.0).

\end{abstract}





\section{Introduction}

Since its discovery in 1995, \object{Gliese 229B} (Gl 229B), a companion to an M dwarf,
has been regarded as the prototype T dwarf \citep{nak95,opp95}.
Although many T dwarfs have been found in the field by large surveys such as SDSS \citep{str99,yor00}, 
2MASS  \citep{bur99,skr06}, UKIDSS \citep{war07,law07} and WISE \citep{mac13,wri10}, Gl 229B remains as one of the best studied T dwarfs to date.
Although the overall energy distribution of Gl 229B was reproduced by early model spectra fairly well \citep{tsu96,mar96,all96},
there were discrepancies in detail between the model and observed spectra.
By these models, the effective temperature was found to be around 900K, but surface gravity and metallicity were poorly known.
According to the progress of observations in late 1990's, attempts were made to improve the models.
\citet{sau00} showed that $(T_{\rm eff}, \log g,$ [M/H]) = (870K, 4.5,-0.5), (940K, 5.0,-0.2) and (1030K, 5.5,0.0) are all acceptable
combinations of the parameters, while \citet{leg02} claimed that the best fit combination of the model parameters is  
$(T_{\rm eff}, \log g,$ [M/H]) = (1000K, 3.5,-0.5) which corresponds to an age of 30 Myr.
Despite the accumulation of more observational data, the determination of stellar parameters is far from conclusive.
In the mean time, more and more T dwarfs were discovered, and
in 2006, a unified near-infrared spectral classification scheme for T dwarfs was presented by
\citet{bur06}. According to this scheme, Gl 229B was classified as a T7 peculiar object, and after this
classification, Gl 229B appears to have been forgotten in the literature.

The metallicity and age of the primary star Gl 229A are expected to be the same as those of Gl 229B.
The metallicity of Gl 229A is controversial and the derived values so far are [M/H]=+0.15$\pm0.15$ \citep{mou78}, 
-0.2$\pm0.4$ \citep{sch97},
and -0.5 \citep{leg02}. The age indicators are the kinematics of the young disk population and stellar activity
as a flare star, and based on them \citet{leg02} claimed that the age of Gl 229A  could be as young as 30 Myr.

In this paper, we apply newly determined carbon and oxygen abundances of Gl 229A \citep{tsu14,tsu15}
to model the atmosphere of Gl 229B. 
We also reexamine the age of Gl 229A from the points of view of the kinematics and stellar activity.
We then give the combinations of the good fit parameters of Gl 229B, which satisfy the age constraint.


\section{Carbon and oxygen abundances of Gl 229A}



The details of the determination of carbon and oxygen  abundances of M dwarfs
including Gl 229A are described elsewhere
\citep{tsu14,tsu15} and
a brief summary is given here.

The $K$-band echelle spectrum of Gl 229A with a resolution of 20,000 was obtained at the Subaru Telescope,
using IRCS \citep{kob00} with the natural guide star adaptive optics.
Before the spectral line analysis, \citet{tsu14} calculated the effective temperature and 
the surface gravity. 
The effective temperature, $T_{\rm eff} = 3710$K was estimated
from the newly derived $\log T_{\rm eff} - M_{3.4}$ relation, where $M_{3.4}$ is 
the absolute magnitude at 3.4 $\mu$m based on the $WISE$ $W1$-band flux and the Hipparcos parallax.
The surface gravity was estimated to be $\log g = 4.77$
using the temperature-radius relation and mass-radius relation derived by \citet{boy12}.

The new analysis of spectral lines is based on the fact that pseudo-continua are seen both on the observed
and model spectra. The observed spectrum of Gl 229A is depressed by numerous weak lines of
H$_2$O, by which a pseudo-continuum is created. On the other hand, both the true- and pseudo-continuum
are evaluated on the model spectrum. The pseudo-continuum on the model spectrum is generated owing to
the recent H$_2$O line database. 
By analyzing EWs of the CO lines affected by H$_2$O contamination on the pseudo-continua of
both observed and model spectra, the difficulty of spectral line analysis of the depressed observed continuum  
can be overcome. 

Almost all the carbon atoms are in the form of CO molecules in the M dwarf atmosphere of Gl 229A.
This situation is not changed by the changes of physical condition in the photosphere and
this is the reason that the numerous CO lines are excellent abundance indicators of carbon.
H$_2$O molecules consume a large portion of oxygen left after the CO formation and they are
stable in the photosphere of Gl 229A. Similarly to the determination of carbon abundance from CO lines,
oxygen abundance is determined from the analysis of H$_2$O lines.

The resultant carbon and oxygen abundances are $\log A_{\rm C}=-3.27\pm0.07$ and $\log A_{\rm O} = -3.10\pm0.02$ respectively.
The carbon abundance is higher than the classical high solar abundance of $\log A_{\rm C} = -3.40$,
while the oxygen abundance is slightly lower than the classical high solar abundance of $\log A_{\rm O} = -3.08$, and 
the C/O ratio of 0.68$\pm$0.12 is slightly higher than the solar value of 0.48 \citep{gre91}.

\section{Brown Dwarf Models of Gl 229B}

We compare the observed spectrum of Gl 229B between 0.8 and 5.1 $\mu$m obtained from the archive by
Leggett (http://staff.gemini.edu/\~{ }sleggett/LTdata.html)
with a set of model spectra based on the carbon and oxygen abundances of Gl 229A.
The Gl 229B spectrum in the archive is a composite of  several spectra with flux calibrations based on photometry:
\citet{sch98} HST/STIS spectrum, retained their calibration,
attached at 1.023 $\mu$m to \citet{geb96} $JHK$ spectrum 
flux calibrated as per \citet{leg99} plus
far-red spectrum fluxed by \citet{gol98}
[HST 0.81 $\mu$m and] 1.04 $\mu$m photometry
includes \citet{opp98} $L$ band spectrum recalibrated 2.98$-$4.15 $\mu$m
and \citet{nol97} 5 $\mu$m spectrum calibrated with UKIRT photometry from \citet{gol04}.

To compare with the observed spectrum, we first use the model spectra in a dust free subset
based on the classical  C \& O abundances (Ca-series based on {\it case a} abundance) from  the
database of  the Unified Cloudy Models
(UCMs: http://www.mtk.ioa.s.u-tokyo.ac.jp/\~{ }ttsuji/export/ucm2)
\citep{tsu02,tsu05}.
Also we newly prepare a small grid of model photospheres (without dust clouds) for the metallicity
scaled to the carbon abundance of Gl 229A, i.e. all the metals  are increased except for He, Li and O by
$-3.27-(-3.40) = +0.13$ dex against the classical solar abundances referred to as {\it case a}.
The oxygen abundance is kept to be $-3.10$, the value for Gl 229A.
We use these abundances to generate model spectra of $R = 600$ (about the same as the resolution of the observed spectrum).
 In the brown dwarf as cool as  Gl 229B, the dust layer lies deep in the atmosphere and
there is not significant effect on the photospheric emission \citep{tsu02}. Therefore
the use of dust-free models is justified. 
 
\begin{figure}
\epsscale{1.0}
\plotone{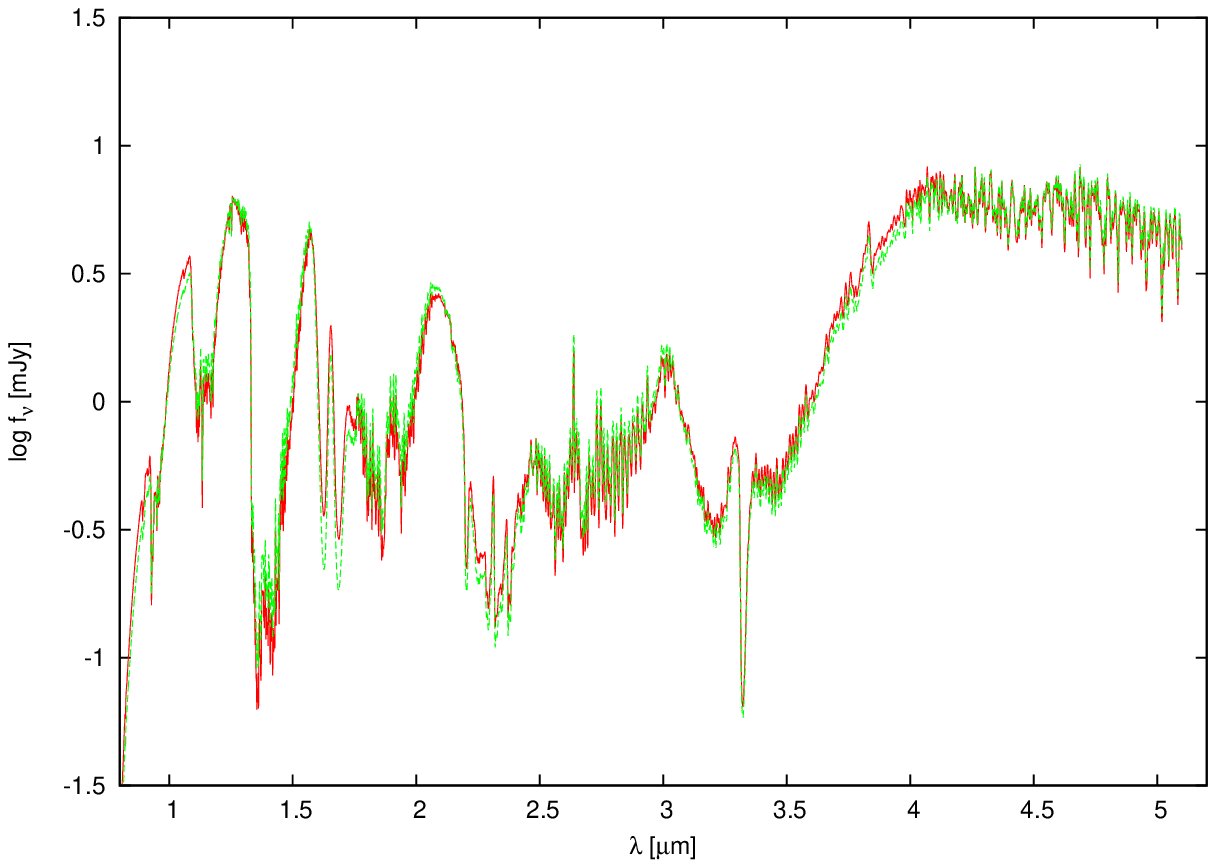}
\caption{Solar metallicity model (Ca-series based on the classical C \& O abundance)
spectrum (red line) and a model spectrum for the metallicity  of Gl 229A (green line) of Gl 229B.
Model parameters are $T_{\rm eff} = 900$K and $\log g =5.0$.
CH$_4$ bands (at 1.6, 2.3 \& 3.5 $\mu$m) are slightly stronger
while H$_2$O bands (at 1.4, 1.8 \& 2.7 $\mu$m) are slightly weaker in the model based on the metallicity of Gl 229A,
as expected for the increased carbon and decreased oxygen. The differences of the spectra, however, are not very
large at the resolution as low as 600.
\label{fig1}}
\end{figure}

Before the comparison of the observed and model spectra, we first investigate the
significance of the effect of the non-solar metallicity of Gl 229A on the Gl 229B's spectrum
by comparing the classical solar metallicity model (i.e. of $\log A_{\rm C} = -3.40$ \& $\log A_{\rm O} = -3.08$)
with the model with the metallicity of Gl 229A
for $T_{\rm eff}$=900K and $\log g$=5.0.
Inspection of Figure \ref{fig1} reveals that CH$_4$ bands (at 1.6, 2.3, \& 3.5 $\mu$m) are slightly stronger while
H$_2$O bands (1.4, 1.8, \& 2.7 $\mu$m)  are slightly weaker in the model based on the metallicity of Gl 229A,
as expected for the increased carbon and decreased oxygen.
The differences of the spectra, however, are not very large at the resolution as low as 600.
In the discussion below, we always use the models
 with the metallicity of Gl 229A, but the differences of the spectra based on the classical solar metallicity and 
 on the metallicity of Gl 229A shown in Figure \ref{fig1}, generally result in the same model parameters, since
 the effect of metallicity difference of 0.1 dex or so on the low resolution spectra is rather modest.
 Thus we confirm that the abundance of Gl 229B is not much different from the classical solar abundance, and
 this is an advantage, since the evolutionary models we use to interpret the spectra  \citep{bur01} 
 are calculated only for the solar metallicity.

The limitations of the model and observed spectra should also be noted before their comparison.
The $H$-band methane absorption in the model spectra is calculated by a band model and the agreement
with the observed spectrum at this portion of the spectrum is not so good. 
The $K$-band methane absorption in the model spectra is calculated by a line list.
However the quality of the line list is not so great and the agreement of the model and observed
spectra at this portion ($2.2-2.4 \mu$m) is not so good either.
The model spectra are based on
equilibrium chemistry and a feature related to disequilibrium chemistry,
the 4.6 $\mu$m CO fundamental band \citep{nol97} is not reproduced.
On the other hand,  the CO absorption feature in the observed spectrum is very noisy
and it cannot be analyzed simultaneously with other features.  
Signal-to-noise ratios at individual data points of the observed spectrum are unknown 
and the $\chi^2$ test based on the absolute value of $\chi^2$ is not possible.

Because of the limitations mentioned above, we focus our attention to good portions of the observed and model spectra,
namely, wavelength segments at ($1.0-1.35\mu$m), ($1.45-1.6$), $(1.8-2.2$) and the red wing of $L$-band methane feature($3.6-4.0\mu$m) for which a line list 
(R. Freedman, private communication) is available for modeling.
Since we do not know the signal-to-noise ratio of each data point, 
as proxy for a $\chi^2$, we define a relative reduced
 $\chi^2$, $\chi^2_r$ by assigning a constant signal-to-noise ratio,
$C=1$ as

\begin{equation}
\chi^2_r = \frac{1}{N-1} \sum_j \left[\frac{C}{f_j(obs)}\left(f_j(obs)-f_j(model)\right)\right]^2,
\end{equation}

where $j$ indicates the $j$th data point and $N$ is the number of data points. 
Since $C = f_j(obs)/\sigma_j$, $C/f_j(obs) = 1/\sigma_j$, where $\sigma_j$ is the standard deviation,
which is unknown like $C$.
There is no significance in the value of $C$. 
Since we are dealing with only one object Gl 229B, the usefulness of the relative reduced $\chi^2_r$ is 
justified.  
There are 1416 data points in all and 
the fitting parameter is the radius of the model brown dwarf.

%
%

A wide range of parameter space is explored in the model fitting.
We calculate 30 model spectra with $T_{\rm eff} = 750, 800, 850. 900. 950$ and 1000K and with $\log g$ = 4.5, 4.75, 5.0, 5.25 and 5.5.
The results of model parameters and goodness of fit, $\chi^2_r$ are given in Table \ref{tbl-1}.
From the radius obtained by fitting and $\log g$ of a model brown dwarf, its mass is derived.

The bolometric luminosity of Gl 229B has three determinations. Combining spectroscopic and photometric data from 0.82 to 10 $\mu$m,
\citet{mat96} found $L = 6.4 \times 10^{-6} L_\odot$. With their own $JHKL^{\prime}$ photometry, \citet{leg99} obtained
 $L = 6.6 \pm 0.6 \times 10^{-6} L_\odot$. From a recalibration using the $HST$ photometry of \citet{gol98}, \citet{sau00} found
$L = 6.2 \pm 0.55 \times 10^{-6} L_\odot$. We here adopt the mid value of the bolometric luminosity  $L = 6.4 \times 10^{-6} L_\odot$ 
and the evolutionary models by \citet{bur01} to estimate the ages of the model brown dwarfs, which are given in 
the column 5 of Table \ref{tbl-1}.

\section{Age of Gl 229A}

\subsection{Kinematics of Gl 229A}

The space velocity of Gl 229A is $(U_\odot,V_\odot,W_\odot) = (+12,-11,-12)$ in units of kms$^{-1}$.
The solar motion in the Local Standard of Rest (LSR) frame of reference is $(U,V,W) = (10.0, 5.2, 7.2)$ \citep{deh98} and
the velocity of Gl 229A in the LSR frame is $v(LSR)=25.5$ kms$^{-1}$, which is significantly 
smaller than the LSR velocity of the average disk population of $v =43$ kms$^{-1}$ \citep{gil99}. 
\citet{leg92} classified this star as a young-disk star, which is reasonable from the value of the LSR velocity.

A quantitative estimate of the age is possible if we assume  that the present LSR velocity of Gl 229A is 
produced by a stochastic process in the galactic disk.
The total velocity dispersion $\sigma_v$ (kms$^{-1}$) in the LSR frame, as a function of stellar age $\tau$ (Gyr), is described by

 \begin{equation}
 \label{sigmav}
 \sigma_v(\tau)^3 = 1000 + 8.19 \times 10^4 [\exp(\tau/T)-1], 
\end{equation}

where $T$ is 5 Gyr \citep{gil99}. 

The fraction of stars with random velocities smaller than $v(LSR)$ for a given velocity dispersion $\sigma$, $F(\sigma)$, is given
by the isotropic Boltzmann distribution,

\begin{equation}
\label{boltzmann}
F(\sigma) = \int_0^{v(LSR)} \sqrt{\frac{2}{\pi}} \frac{1}{\sigma^3} \exp\left({-\frac{v^2}{\sigma^2}}\right) v^2 dv.
\end{equation}

The values of $F(\sigma)$ are given for $\tau$ and $\sigma$ in Table \ref{tbl-2}. 
According to the table, the probability that
Gl 229A is actually younger than 3.0 Gyr is $1-F(\sigma)$ = 94\%. We adopt 3.0 Gyr as the upper limit to the age of Gl 229A.

\subsection{Stellar activity of Gl 229A}

While Gl 229A is shown to flare, the time-integrated energy in flare light is very much lower than for `normally active' flare stars.
M dwarfs with high chromospheric activity show Balmer lines in emission, while Gl 229A shows Balmer lines in absorption indicating a low level of chromospheric activity \citep{byr85}. The stellar activity indicates that Gl 229A is not extremely young, but not so old.
As a quantitative analysis, 
\citet{giz02} obtained a lower limit to the age of Gl 229A, based on the absence of H$\alpha$ emission and their argument is
summarized in the following. 

The analysis of M dwarfs in open clusters by \citet{haw99} suggests an approach to an age-activity relation. While the activity 
levels of stars in a given cluster exhibit considerable scatter, there is a well-defined $V-I_C$ color at which activity becomes ubiquitous.
All stars redder than this color are dMe (defined as EW(H$\alpha$)$>1.0$\AA), while the bluer stars are dM without emission. 
Observations exist for M dwarfs in six clusters: IC2602 and IC2391(30 Myr), NGC 2516 and Pleiades (125 Myr), Hyades (625 Myr), and
M67 (4.0 Gyr). \citet{haw99} have used these observations to determine the relationship between the ``H$\alpha$ limit'' color and the age,

\begin{equation}
V - I_C = 2.54 + 1.05 \log \tau,
\end{equation}

where $\tau$ is the age in Gyr.
For Gl 229A, \citet{giz02} used H$\alpha$ absorption and $V-I=2.01$ to derive the lower limit to the age, 0.32 Gyr.
Based on this age limit, Gizis et al. rule out the young age of 30 Myr advocated by \citet{leg02} and suggest that the synthetic spectra
from model atmospheres are not yet adequate for age determinations.
We adopt 0.3 Gyr as the lower limit to the age of Gl 229A, taking into account the error in the use of $V-I$ instead of $V-I_C$.

\section{Interpretation of Models of Gl 229B}

From the previous section, we have given the range of the age of Gl 229A to be between 0.3 and 3.0 Gyr.
In the column 7 of Table \ref{tbl-1}, whether a model satisfies this age constraint or not is given by Y or N.
The highest gravity models for $\log g = 5.5$ are all ruled out, because the model brown dwarfs
for $T_{\rm eff}\le900$K are more massive
than the hydrogen burning limit and the model brown dwarfs for $T_{\rm eff}=950$K and 1000K
are older than 3 Gyr. 
The relative reduced $\chi^2$,
$\chi^2_r$ is given  in the column 6  and the ranking of
the goodness of fit(GOF) is given by A$(\chi^2_r<0.070)$, B$(0.070\le\chi^2_r<0.075)$,
and C$(0.075\le\chi^2_r)$ in the column 8.

\begin{figure}
\epsscale{1.0}
\plotone{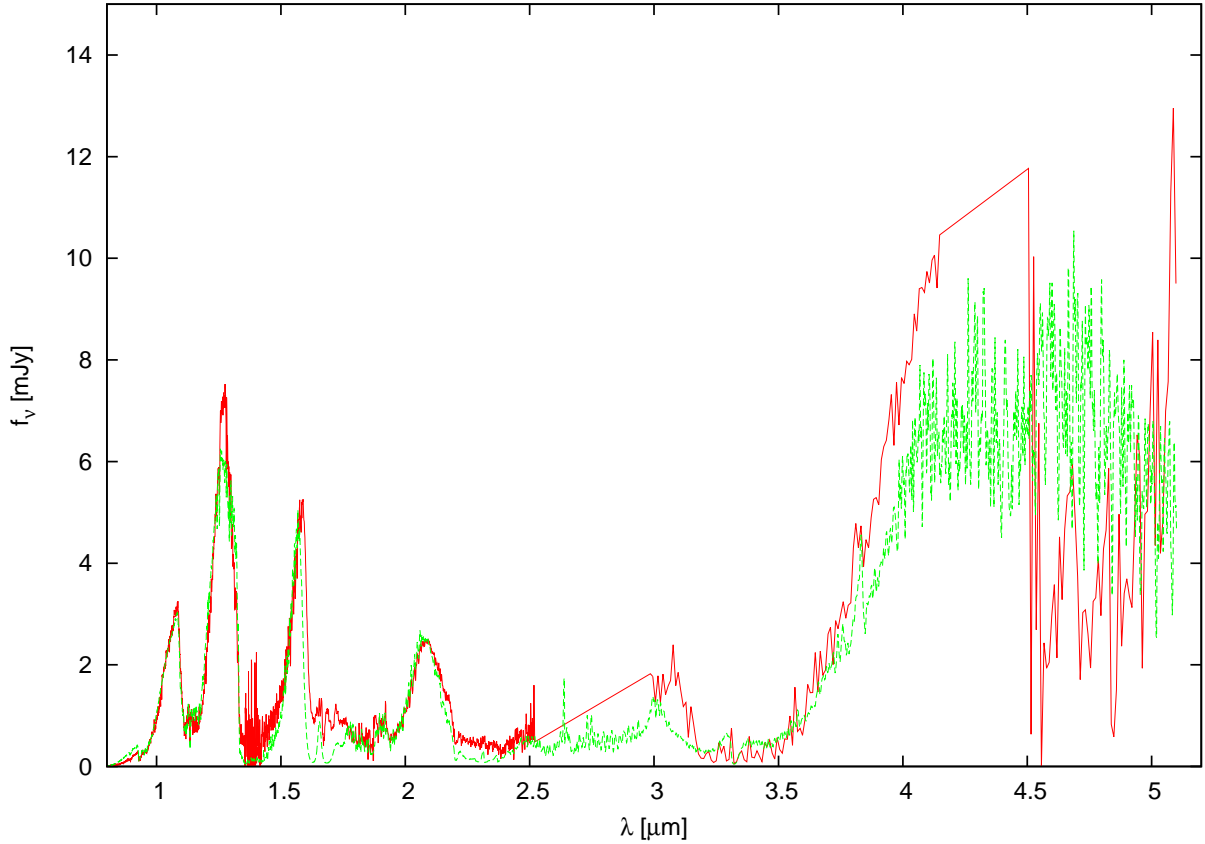}
\caption{Observed spectrum (red line) of Gl 229B and 
a model spectrum for the metallicity of Gl 229A (green line) .
Model parameters are ($T_{\rm eff}, \log g, m, t)$ = (800K,4.75,0.028$M_\odot$,1.1Gyr). 
This is one of the two models with the ranking of goodness of fit GOF=A
that satisfy the age constraint.
There are two gaps in the observed spectrum between 2.59 and 2.99 $\mu$m and between 4.15 and 4.51 $\mu$m.
\label{fig2}}
\end{figure}

There are  two models which satisfy the age constraint(Y), and GOF=A, whose physical parameters are 
$(T_{\rm eff}, \log g, m, t)$ = (800K,4.75,0.028$M_\odot$,1.1Gyr)
and (850K,5.0,0.037$M_\odot$,2.3Gyr).
These are the best fit models of all, and the spectrum for the former model is plotted in Figure \ref{fig2}.

\begin{figure}
\epsscale{1.0}
\plotone{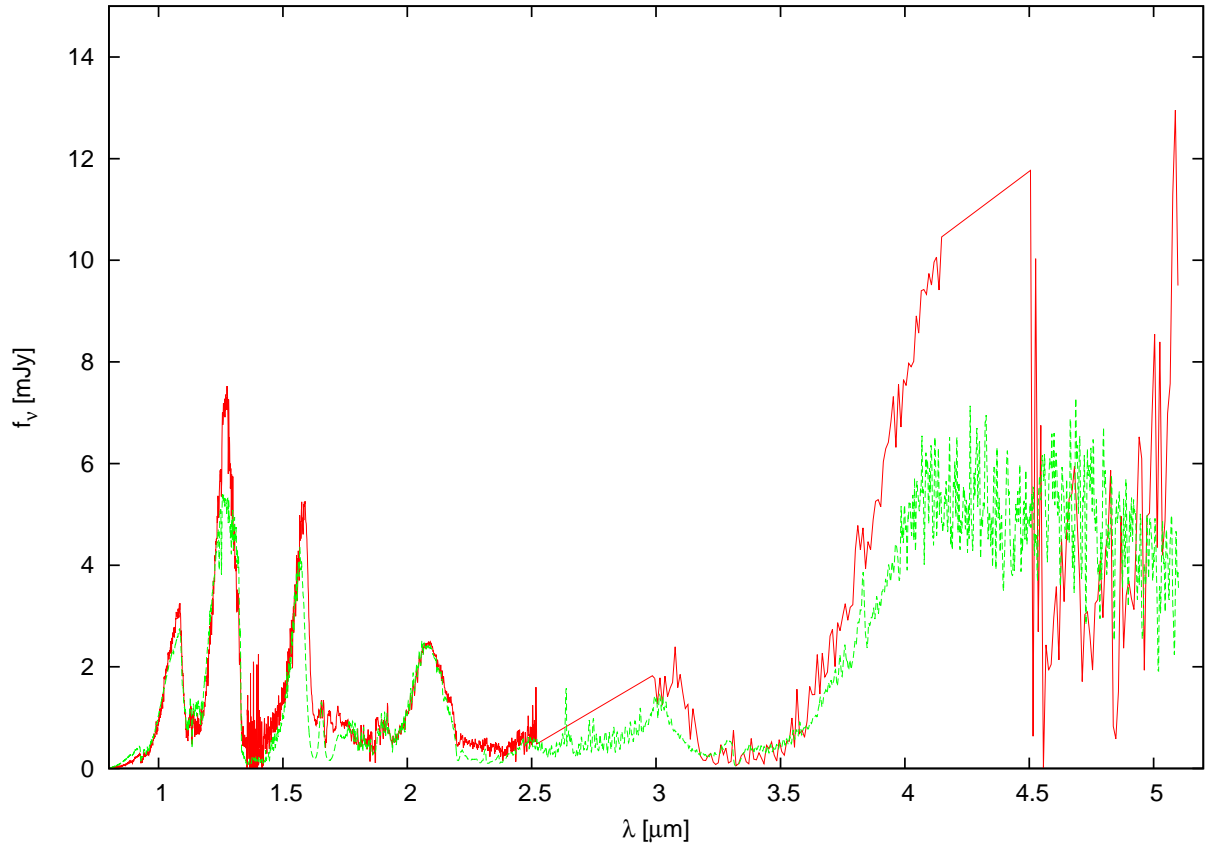}
\caption{Observed spectrum (red line)  of Gl 229B and 
a model spectrum for the metallicity of Gl 229A (green line) .
Model parameters are ($T_{\rm eff}, \log g, m, t)$ = (900K, 5.0, 0.028$M_\odot$, 1.1Gyr). 
This is one of the three models with the ranking of goodness of fit GOF=B that 
satisfy the age constraint.  These models show still reasonable fit.
\label{fig3}}
\end{figure}

There are three models with (age constraint, GOF)=(Y,B).
Their physical parameters are 
$(T_{\rm eff},\log g, m, t)$ = (750K,4.75,0.038$M_\odot$,2.5Gyr),(850K,4.75,0.021$M_\odot$,0.70Gyr), 
and (900K,5.0,0.028$M_\odot$,1.1Gyr).
We consider that these models are still good fit to the observation.
As an example, the spectrum for the parameters, (900K,5.0,0.028$M_\odot$,1.1Gyr)  is plotted in Figure 
\ref{fig3}.

\begin{figure}
\epsscale{1.0}
\plotone{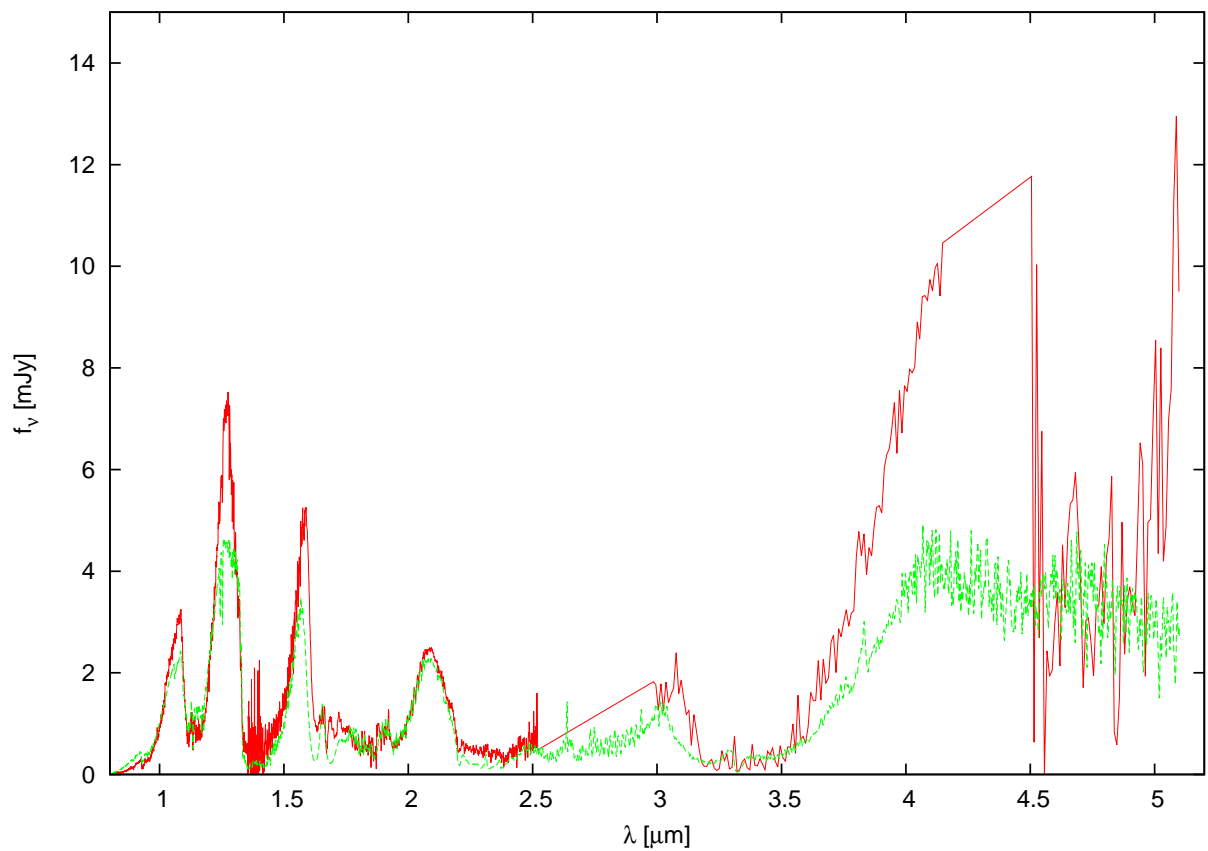}
\caption{Observed spectrum (red line) of Gl 229B and 
a model spectrum for the metallicity of Gl 229A (green line) .
Model parameters are ($T_{\rm eff}, \log g, m, t)$ = (1000K, 5.0, 0.016$M_\odot$, 0.48Gyr). 
This is one of the seven models with the ranking of goodness of fit  GOF=C  
that satisfy the age constraint. We regard that the fits are poor for these models.
\label{fig4}}
\end{figure}

The rest of the seven models which satisfy the age constraint are ranked C. There is no
(Y,A) or (Y,B) models for $T_{\rm eff}=$950 and 1000K. The high temperature models apparently
fit poorly compared to the models for lower temperatures. 
As an example of the poor fit models, the spectrum for parameters,(1000K,5.0,0.016$M_\odot$,0.48Gyr),
is plotted in Figure \ref{fig4} 

To summarize,
we consider that the spectra for (Y,A) models 
are the best fits and those for (Y,B) models are reasonable fits to the observed spectrum.
In terms of the effective temperature and surface gravity, 
($T_{\rm eff},\log g$)=(800K,4.75) and (850K,5.0) are the best fit parameters,
while 
(750K,4.75),(850K,4.75) and (900K,5.0)
are still acceptable model parameters.


\section{Comparison with Other Works} 

\subsection{Metallicity of Gl 229A} 

As far as we are aware, there have been three determinations of the metallicity of Gl 229A. 
\citet{mou78} obtained high resolution spectra of $H$ and $K$ band using the Fourier Transform Spectrometer
at KPNO with a resolution $R \sim 20,000$. He analyzed Al, Ca, and Mg lines and found [M/H]=+0.15$\pm0.15$.
The effective temperature and the surface gravity he adopted are  $T_{\rm eff} =3614$K and $\log g =4.75$ respectively,
which are in good agreement with the estimates  $T_{\rm eff} =3710$K and $\log g =4.77$ by \citet{tsu14}. 

\citet{sch97} obtained [Fe/H]=-0.2$\pm 0.4$ by the analysis of the FeH Wing-Ford band at around 1 $\mu$m from 
a medium resolution spectrum of $R \sim 12,500$. The effective temperature of their best fit model of $T_{\rm eff} =3300$K
is significantly lower than the estimate by \citet{tsu14}.
The two determinations of the metallicity  \citep{mou78,sch97} are both consistent with solar metallicity
(within the error bars).

\citet{leg02} fit with model atmospheres a spectrum covering from 0.8 to 2.5 $\mu$m with a resolution $R\sim 800$
and a spectrum covering from 1.12 to 1.22 $\mu$m with a resolution $R\sim 5,800$. 
Best fit model parameters are $(T_{\rm eff}, \log g$, [M/H]) = (3700K, 4.0, -0.5) and (3700K, 3.5, -0.7).
Although the effective temperature is in good agreement with our estimate, $\log g$ and [M/H] are significantly
lower than our values $\log g = 4.77$ and [C/H] = +0.13$\pm0.07$. 
The metallicity obtained by \citet{leg02} is definitely lower than solar metallicity.
The application of evolutionary models implies that
the age of Gl 229A is as young as 30 Myr. 
As mentioned in Section 4.2, \citet{giz02} rule out this young age and suggest that the problem is in the model spectra
by \citet{all01} used by \citet{leg02}.

\subsection{Metallicity of Gl 229B}


\citet{sau00} do not use any information about Gl 229A and  their abundance analysis is based on the bolometric luminosity
$L$ of Gl 229B alone. Since $L$ is known, a determination of the surface gravity $g$ fixes
$T_{\rm eff}$, the radius, the mass, and the age of Gl 229B, as well as the metallicity. 
It is assumed that evolutionary models for solar metallicity \citep{bur97}  are applicable for non solar metallicity.
They fit the observed and model spectra by eye.
They analyze the ``red'' spectrum from 0.83 to 1 $\mu$m ($R\sim 2250$), $J  (R\sim 2400)$, $H (R\sim 2100)$, 
and $K (R\sim 2800)$ spectra  separately, and derive the H$_2$O abundances, or [O/H]s. 
They interpret [O/H] as the metallicity [M/H]. 
No over-all fitting from 0.8 to 5 $\mu$m is given.
For a given $\log g$, the optimal [M/H] was obtained out of a set of four [M/H]s for different wavelength bands.
For three values of $\log g$, they generated three models (ABC) with optimal parameters: ($T_{\rm eff}$, $\log g$, [M/H]) =
A(870K, 4.5,-0.5), B(940K,5.0,-0.3) and C(1030K, 5.5,-0.1). They quote the error in abundance to be $\pm 0.1$.
In terms of the metallicity the model C is the closest to our models, but $\log g$ is out of the range. 
The compromise is the model B for which the counterpart of our model is  ($T_{\rm eff}$, $\log g$, [M/H]) =(900K,5.0,+0.13).

\citet{leg02} use the 0.84$-$4.15 $\mu$m spectra for Gl 229B taken from \citet{leg99} and also use
the 4.5$-$5.1 $\mu$m spectrum from \citet{nol97} to fit with the model spectra by eye.
The best fit model parameters are ($T_{\rm eff},$, $\log g$. [M/H]) = (1000K, 3.5, -0.5).
Although the metallicity [M/H]=-0.5 is in agreement with the value obtained by \citet{sau00}, the surface gravity
$\log g$=3.5 is extremely small.  
The inspection of the best fit model spectra in comparison with the observed spectrum
reveals significant discrepancies in $Z$, $H$, and $K$ bands. 
These discrepancies are much greater than those seen in our fitting of $ZJHK$ bands. 
The H$_2$O line list used by \citet{all01} is of \citet{par97}, while that by us is of the BT2-HITEMP2010 database \citep{bar06,rot10}.
However, the effect of line lists may be minor at low resolution.
Higher resolution spectra at $J$, $H$, and $K$ bands \citep{sau00} are also fitted by \citet{leg02} with the best fit model spectra
obtained from the analysis at the lower resolution.
Although the H$_2$O line list used by \citet{sau00} was the same as that by \citet{all01}, the fit by \citet{sau00} is
significantly better especially at $H$ and $K$ bands.
We suspect that the $\log g = 3.5$ adopted by \citet{leg02}
is not adequate and consider that the range of $\log g = 5.0 \pm 0.5$ adopted by \citet{sau00} is more appropriate.

\section{Disequilibrium Chemistry}

Our analysis is based on equilibrium chemistry and we do not analyze the 4.6 $\mu$m CO fundamental band.
However, the effects of disequilibrium chemistry can be important for T dwarfs \citep{nol97,opp98,gri99,sau00,vis11,zah14}
and we briefly mention here the effect of the Gl 229B metallicity on the CO abundance estimation. 
In general, higher gravities, cooler temperatures and lower metallicity favor CH$_4$ vs. CO.
Some workers estimate CO abundances, assuming low metallicity ([M/H]$\le-0.3$) \citep{nol97,vis11}, while
others give CO abundances for low ([M/H]$\le -0.5$) and solar metallicity \citep{gri99,sau00}.
Since the carbon abundance we adopt is [C/H] = +0.13 \citep{tsu14}, CO abundances derived for solar metallicity
are of our interest.
\citet{gri99} estimate $\log X_{\rm CO} \ge -4$ for [M/H]=0.0, while \citet{sau00} give $\log X_{\rm CO} = -3.5$ for [M/H]=$-0.1$.
These values are significantly higher than for example $-4.3 \le \log X_{\rm CO} \le -3.7$ obtained by \citet{nol97}.

 


 \section{Concluding remark}
 

It has been two decades, since the discovery of Gl 229B as the first genuine brown dwarf in 1995.
It had been the only known T dwarf, before a small number of field T dwarfs were discovered in 1999.
As the prototype T dwarf, Gl 229B was studied intensively until early this century and efforts were
made to determine its physical properties. There are two approaches in the determination of
physical properties; one is to derive $T_{\rm eff}$, $\log g$ and [M/H] all from the model fitting
of the Gl 229B spectrum itself and the other is to derive $T_{\rm eff}$ and $\log g$ from
the model fitting of the Gl 229B spectrum, but use [M/H] obtained from
the spectrum of the primary star, Gl 229A.
Although the latter approach is favorable because of the fewer parameters in the model fitting of
the Gl 229B spectrum,
it was deemed difficult because Gl 229A is an M dwarf.  
It was not until recently that the metallicity (abundances of C and O) was determined reliably.
We used these newly determined C and O abundances of Gl 229A in the model fitting of the Gl 229B spectrum. 
In addition to the metallicity, the ages of the binary components are safely assumed to be the same.
We estimated the range of the age of Gl 229A based on its activity and kinematics. 
This age constraint narrowed the combinations of physical parameters of acceptable models
using the evolutionary models of brown dwarfs.   
It took twenty years to settle the question of physical properties of Gl 229B, although
early guesses assuming the solar metallicity were not so bad.

\acknowledgments

We thank T. Geballe and S. Leggett for providing us with the Gl 229B spectrum.
We also thank the anonymous referee for helpful comments which significantly improved
the manuscript.

\clearpage



\clearpage

\clearpage

\clearpage

\clearpage

\begin{deluxetable}{cccccccc}
\tabletypesize{\scriptsize}
\tablecaption{Model Parameters and goodness of fit\label{tbl-1}}
\tablewidth{0pt}
\tablehead{
\colhead{$T_{\rm eff}$} & \colhead{$\log g$} & \colhead{$r$} & \colhead{$m$} & \colhead{$t$} &
\colhead{$\chi^2_r$\tablenotemark{a}} &
\colhead{$0.3\leq t \leq 3.0$} & \colhead{GOF\tablenotemark{b}} \\ 
\colhead{K}    &      & \colhead{$R_\odot$} & \colhead{$M_\odot$} & \colhead{Gyr}  & & &
}
\startdata
750  & 4.50 & 0.136 & 0.021 & 0.70 & 0.0764 & Y & C \\
     & 4.75 & 0.137 & 0.038 & 2.5  & 0.0716 & Y & B \\
     & 5.00 & 0.137 & 0.068 & $>$10& 0.0738 & N & B \\
     & 5.25 & 0.137 & 0.122 & ---  & 0.0781 & N & C \\
     & 5.50 & 0.137 & 0.216 & ---  & 0.0903 & N & C \\
800  & 4.50 & 0.116 & 0.016 & 0.47 & 0.0752 & Y & C \\
     & 4.75 & 0.117 & 0.028 & 1.1  & 0.0696 & Y & A  \\
     & 5.00 & 0.118 & 0.051 & 4.3  & 0.0673 & N & A  \\
     & 5.25 & 0.118 & 0.090 & ---  & 0.0685 & N & A   \\
     & 5.50 & 0.118 & 0.161 & ---  & 0.0748 & N & B  \\
850  & 4.50 & 0.099 & 0.011 & 0.20 & 0.0802 & N & C  \\
     & 4.75 & 0.100 & 0.021 & 0.70 & 0.0730 & Y & B  \\
     & 5.00 & 0.101 & 0.037 & 2.3  & 0.0682 & Y & A  \\
     & 5.25 & 0.101 & 0.066 & $>$10& 0.0667 & N &  A  \\
     & 5.50 & 0.102 & 0.120 & ---  & 0.0705 & N &  B  \\
 900 & 4.50 & 0.086 & 0.009 & 0.10 & 0.0889 & N &  C  \\     
     & 4.75 & 0.087 & 0.016 & 0.47 & 0.0796 & Y &  C  \\
     & 5.00 & 0.088 & 0.028 & 1.1  & 0.0732 & Y &  B  \\   
     & 5.25 & 0.088 & 0.050 & 4.3  & 0.0692 & N & A   \\
     & 5.50 & 0.089 & 0.091 & ---  & 0.0691 & N & A  \\
 950 & 4.50 & 0.075 & 0.006 & 0.06 & 0.1010 & N & C  \\
     & 4.75 & 0.076 & 0.012 & 0.24 & 0.0900 & N & C  \\
     & 5.00 & 0.077 & 0.022 & 0.75 & 0.0817 & Y & C  \\
     & 5.25 & 0.077 & 0.038 & 2.5  & 0.0764 & Y & C  \\
     & 5.50 & 0.078 & 0.070 & $>$10& 0.0739 & N & C  \\ 
 1000& 4.50 & 0.066 & 0.005 & 0.04 & 0.1145 & N & C  \\ 
     & 4.75 & 0.067 & 0.009 & 0.13 & 0.1019 & N & C  \\
     & 5.00 & 0.067 & 0.016 & 0.48 & 0.0935 & Y & C  \\
     & 5.25 & 0.068 & 0.030 & 1.5  & 0.0846 & Y & C  \\
     & 5.50 & 0.069 & 0.055 & 4.9  & 0.0803 & N & C  \\           
\enddata
%
\tablenotetext{a}{Relative reduced $\chi^2$}
\tablenotetext{b}{Ranking of goodness of fit based on relative reduced $\chi^2$. A($<$0.07), B(0.07$\sim$0.075), and C($>$0.075).}
\end{deluxetable}



\clearpage

\begin{table}
\begin{center}
\caption{Age, velocity dispersion and fraction of stars with $v<v(LSR)$ \label{tbl-2}}
\begin{tabular}{ccc}
\tableline\tableline
$\tau$   & $\sigma$  & $F(\sigma)$\tablenotemark{a} \\
 Gyr             &  kms$^{-1}$             &                  \\
\tableline
0.5 & 21 & 0.312 \\
1.0 & 27 & 0.172 \\
1.5 & 31 & 0.121 \\
2.0 & 35 & 0.088 \\
3.0 & 41 & 0.057 \\
average disk & 43 & 0.050 \\
%
%
\tableline
\end{tabular}
\tablenotetext{a}{$F(\sigma)$ is the fraction of stars whose velocities 
generated by a stochastic process are smaller than $v(LSR)$
The probability that a star is younger than the age $\tau$ is given by
$1-F(\sigma)$.}
\end{center}
\end{table}



\end{document}